\documentclass{pasa}%

\usepackage{graphicx}
\usepackage{longtable}

\title[$\gamma$-ray Doppler factor]{The estimation of $\gamma$-ray Doppler factor for {\it Fermi}/LAT-detected blazars}

\author[Pei et al.]{Zhiyuan Pei$^{1,2,3,4,5,6,7}$, Junhui Fan$^{3,4,5,6,7}$\thanks{Corresponding author: fjh@gzhu.edu.cn}, Jianghe Yang$^{3,8}$ and Denis Bastieri$^{1,2,3}$
\affil{$^1$Dipartimento di Fisica e Astronomia ``G.\ Galilei'', Universit\`a di Padova, I-35131 Padova, Italy}
\affil{$^2$Istituto Nazionale di Fisica Nucleare, Sezione di Padova, I-35131 Padova, Italy}
\affil{$^3$Center for Astrophysics, Guangzhou University, Guangzhou 510006, China}
\affil{$^4$Astronomy Science and Technology Research Laboratory of Department of Education of Guangdong Province, Guangzhou 510006, China}
\affil{$^5$Department of Physics and Electronics Science, Hunan University of Arts and Science, Changde 415000, China}
\affil{$^6$Astronomy Science and Technology Research Laboratory of Department of Education of Guangdong Province, Guangzhou 510006, China}
\affil{$^7$Key Laboratory for Astronomical Observation and Technology of Guangzhou, Guangzhou 510006, China}
\affil{$^8$Department of Physics and Electronics Science, Hunan University of Arts and Science, Changde 415000, China}
}

\jid{PASA}
\doi{10.1017/pas.\the\year.xxx}
\jyear{\the\year}

\usepackage{aas_macros}
\usepackage{hyperref} 
\hypersetup{colorlinks,citecolor=blue,linkcolor=blue,urlcolor=blue}

\hypersetup{draft}
\begin{document}

\begin{frontmatter}
\maketitle

\begin{abstract}
Blazars are a subclass of active galactic nuclei (AGNs) with extreme observation properties, which is caused by the beaming effect, expressed by a Doppler factor ($\delta$), in a relativistic jet. Doppler factor is an important parameter in the blazars paradigm to indicate all of the observation properties, and many methods were proposed to estimate its value. In this paper, we present a method following Mattox et al. to calculate the lower limit on $\gamma$-ray Doppler factor ($\delta_{\gamma}$) for 809 selected {\it Fermi}/LAT-detected $\gamma$-ray blazars by adopting the available $\gamma$-ray and X-ray data. Our sample included 342 flat-spectrum radio quasars (FSRQs) and 467 BL Lac objects (BL Lacs), out of which 507 sources are compiled with available radio core-dominance parameter ($R$) from our previous study. Our calculation shows that the average values of the lower limit on $\delta_{\gamma}$ for FSRQs and BL Lacs are $\left<\delta_{\gamma}|_{\rm FSRQ}\right> = 6.87 \pm 4.07$ and $\left<\delta_{\gamma}|_{\rm BL \, Lac}\right>=4.31 \pm 2.97$, respectively. We compare and discuss our results with those from the literature. We found that the derived lower limit on $\delta_{\gamma}$ for some sources are higher than that from the radio estimation, which could be possibly explained by the jet bending within those blazars. Our results also suggest that the $\gamma$-ray and radio regions perhaps share the same relativistic effects. The $\gamma$-ray Doppler factor has been found to be correlated with both the $\gamma$-ray luminosity and core-dominance parameter, implying that the jet is possibly continuous in the $\gamma$-ray bands, and $R$ is perhaps an indicator for a beaming effect. 
\end{abstract}

\begin{keywords}
galaxies: active -- (galaxies:) BL Lacertae objects: general -- (galaxies:) quasars: general -- gamma rays: general -- galaxies: jets
\end{keywords}
\end{frontmatter}

\section{INTRODUCTION }
\label{sec:intro}

Active galactic nuclei (AGNs), the most luminous objects with supermassive black holes lurking at their centers, are among the most energetic sources in the Universe and play a crucial role in the evolution of galaxies. Blazars, as the most extreme subclass of AGNs with a radio-loud behaviour and a relativistic jet pointing toward the observer \citep{UP95}, are characterized by having large amplitude and rapid variability, superluminal motion, high polarization, core-dominated nonthermal continuum, and $\gamma$-ray emission, etc \citep{Wil92, FX96, Bai98, Rom02, Fan05, Fan11, Fan13(b), Fan16, Ghi10, Abd09, Abd10c, Abd10d, Urr11, Mar11, Nol12, Yan12, Yan19, Gup12, Ace15, Xia19, LAT19(b), Pei19, Pei20(b), Pei20(a)}. All of these properties are due to the relativistic beaming effect. The emissions in the jet are highly boosted along the line of the observer's sight. The spectral energy distribution (SED) of the broad-band continuum emission (from radio to $\gamma$-ray) of blazars are usually dominated by two spectral bumps. The low energy bump, from radio through optical/UV (X-rays, in some cases), ascribes to the synchrotron emission from the relativistic electrons in the jet. The second bump, located in the high-energy (X-ray through $\gamma$-ray), is believed to be emanated from the inverse Compton scattering of low energy photons. According to the optical spectral features, blazars are grouped into flat-spectrum radio quasars (FSRQs) and BL Lac objects (BL Lacs) \citep{SF97}. A more physical classification between FSRQs and BL Lacs can be distinguished via their SED synchrotron peak frequencies $\log \, \nu_{p}$. Low-synchrotron peaked blazars (LSP) are characterized by $\log \, \nu_{p}$(Hz)<14, and intermediate-synchrotron peaked blazars (ISP) have 14<$\log \, \nu_{p}$(Hz)<15, while $\log \, \nu_{p}$(Hz)>15 pertains to high-synchrotron peaked blazars (HSP). The majority of ISP and HSP blazars have been classified as BL Lacs, while LSP ones include FSRQs and some low-frequency-peaked BL Lacs \citep[see][and references therein]{Abd10d, Fan16, Bot19}. 

Based on a relativistic beaming model, \citet{US84} proposed that the total emission from AGNs is from two components, namely, a beamed component (core component) and an unbeamed one (extended component). Then, the observed total luminosity, $L^{\rm tot}$, is the sum of the beamed, $L_{\rm b}$, and unbeamed, $L_{\rm unb}$ emissions, i.e. $L^{\rm tot}=L_{\rm b}+L_{\rm unb}$. In the radio band, the ratio of the two components, $R$, is defined as the core-dominance parameter, i.e. $R=L_{\rm b}/L_{\rm unb}$ \citep[see][and references therein]{OB82, Fan11, Pei16, Pei19, Pei20(b), Pei20(a)}, and also can be expressed as $R=S_{\rm core}/S_{\rm ext}$, where $S_{\rm core}$ and $S_{\rm ext}$ refer to the flux density derived from the core and the extended components of radio emission. 

In addition, due to the relativistic beaming effect, the emissions from the jet are strongly boosted in the observer's frame, i.e. $S^{\rm ob}=\delta^{p}S^{\rm in}$, where $S^{\rm ob}$ is the observed emission, $S^{\rm in}$ is the intrinsic emission in the source frame, and $\delta$ is the Doppler factor. The value of $p$ hinges on the physical detail of the jet and geometrical shape of the emitted spectrum \citep{LB85}, $p=2+\alpha$ for continuous jet while $p=3+\alpha$ for a moving compact source, $\alpha$ is the spectral index ($f_{\nu}\propto \nu^{-\alpha}$). 

The Doppler boosting factor can be expressed by $\delta=[\Gamma(1-\beta\cos\theta)]^{-1}$, where $\Gamma$ is a Lorentz factor ($\Gamma=1/\sqrt{1-\beta^{2}}$), $\beta$ is the jet speed in units of the speed of light and $\theta$ is the viewing angle between the jet and the line-of-sight. The Doppler factor is a crucial parameter in the jet of blazars since it reckons how strongly the flux densities are boosted and timescales compressed in the observer's frame. However, it is difficult for us to determine this parameter since it is unobservable. Therefore, some feasible methods have been proposed \citep{Ghi93, Mat93, LV99, Fan09, Hov09, Lio18, Zha20}.   

From the previous studies, the core-dominance parameter $R$ can take the role of the indicator of Doppler boosted beaming effect \citep[see][]{UP95, Fan03},
\begin{equation}
R = f_{\rm in} \Gamma^{-n}[(1 - \beta \cos \theta)^{-n + \alpha} + (1 + \beta \cos \theta)^{-n + \alpha}],
\label{eq1}
\end{equation}
where $f_{\rm in}$ is a ratio, defined by the intrinsic flux density in the jet to the extended flux density in the co-moving frame, $f_{\rm in} = \frac{S^{\rm in}_{\rm core}}{S^{\rm in}_{\rm ext.}}$, $\alpha$ is the spectral index and $n$=2 or 3.

After the launch of {\it Fermi} Large Area Telescope (hereafter, {\it Fermi}/LAT), many new high-energy $\gamma$-ray sources were detected, revolutionising, in particular, the knowledge of $\gamma$-ray blazars, providing us with the opportunity to study the $\gamma$-ray production mechanism. Based on the first eight years of data from the {\it Fermi} Gamma-ray Space Telescope mission, the latest catalog, 4FGL, or the fourth {\it Fermi} Large Area Telescope catalog of high-energy $\gamma$-ray sources, has been released, which includes 5098 sources above the significance of $4\sigma$, covering the 50 MeV$-$1 TeV range \citep{LAT19(a),LAT19(b)}, about 2000 more than the previous 3FGL catalog \citep{Ace15}.  AGNs are the vast majority of sources in 4FGL; among them 2938 blazars, or 681 FSRQs, 1102 BL Lacs and 1152 blazar candidates of unknown class \citep[BCUs,][]{LAT19(a)}.

The previous studies have probed the correlation between $\gamma$-ray emission and radio emission for the selected $\gamma$-ray loud blazars and showed that the $\gamma$-ray emission is strongly beamed \citep{DG95, Fan98}. Consequently, the $\gamma$-ray Doppler factor ($\delta_{\gamma}$) can be estimated for each $\gamma$-ray loud blazars accordingly \citep{Fan99, Fan05}. 

\citet{Pei19} had compiled a catalog listing 2400 AGNs with available core-dominance parameters ($\log \, R$), 770 of which are blazars. It was found that blazars have, on average, higher $\log \, R$ than those non-blazars objects, indicating that blazars are more core dominated \citep[see also][]{Fan11}. \citet{Pei20(a)} analysed a larger sample of 4388 AGNs with available $\log \, R$, 584 are {\it Fermi}/LAT-detected blazars from 4FGL, and obtained that the $\langle\log R\rangle$  for {\it Fermi} blazars is higher than that for non-{\it Fermi}-detected blazars. This is the evidence that the $\gamma$-ray emission is strongly beamed \citep{Ghi93, DG95, Fan13(b), Pei16}.  

In this paper, we estimate the lower limit on $\gamma$-ray Doppler factors for those $\gamma$-ray blazars following \citet{Mat93} as did in \citet{Fan13(a), Fan14}, probing their relations and shedding new light on the relativistic beaming effect of $\gamma$-ray loud blazars. The methodology is discussed in Section 2, while in Section 3 we describe the sample and results. In Section 4 we present the statistical analysis and make the discussion. Finally we draw the conclusions in Section 5. Throughout this paper, we apply the $\Lambda$CDM model, with $\Omega_{\Lambda} \simeq 0.73$, $\Omega_{M} \simeq 0.27$, and $H_{0} \simeq 73 \rm {km \cdot s^{-1} \cdot Mpc^{-1}}$.

\section{METHODOLOGY}

The extreme observation properties of blazars, e.g. rapid variability, high $\gamma$-ray luminosity, core emission dominated, and superluminal motion, is believed to be in connection with the relativistic beaming model. The high-energy $\gamma$-ray emission detected from blazars indicate that the $\gamma$-rays should be strongly beamed, otherwise the $\gamma$-rays would have been absorbed by the lower-energy photons due to pair-production in the collision. Following the idea of \citet{Mat93}, and as did in \citet{Fan13(a), Fan14}, we assume that:

(i) X-ray is produced in the same region as $\gamma$-ray, and the intensities of X-ray and $\gamma$-ray are semblable when $\gamma$-ray emission is observed;

(ii) the emission region is spherical;

(iii) the emission is isotropic, and the size of the emission region is constrained by the timescale of variability, $\Delta T$, to be less than $R_{\rm size}=c\delta\Delta T/(1+z)$, where c is the speed of light, $\delta$ is the Doppler factor and $z$ denotes the redshift, we derive the optical depth for the pair-production \citep{Mat93}, 
\begin{equation}
\begin{aligned}
\tau=2\times10^{3}\left[(1+z)/\delta\right]^{4+2\alpha}\left(1+z-\sqrt{1+z}\right)^{2}h_{75}^{-2}\Delta T_{5}^{-1} \\
\times\displaystyle\frac{F_{\rm 1\,keV}}{\mu \rm Jy}\left(\frac{E_{\gamma}}{\rm GeV}\right)^{\alpha},
\end{aligned}
\label{eq2}
\end{equation}
where $\alpha$ is the X-ray spectral index ($F_{\nu X}\propto\nu_{X}^{-\alpha}$), $h_{75}=$H$_{0}$/75, $\Delta T_{5}=\Delta T/(10^{5})$s, $\Delta T$ is the timescale in units of hour, $F_{\rm 1\,KeV}$ is the flux density at 1 KeV in units of $\mu$Jy and $E_{\gamma}$ denotes the $\gamma$-ray photon energy in units of GeV.                
As the luminosity distance in units of Mpc can be expressed in the form
\begin{equation}
d_{L}=\displaystyle\frac{c}{H_{o}}\int^{1+z}_{1}\frac{dx}{\sqrt{\Omega_{M}x^{3}+1-\Omega_{M}}},
\label{eq3}
\end{equation}
then the optical depth $\tau$ can be rewritten into 
\begin{equation}
\begin{aligned}
\tau=1.54\times10^{-3}\displaystyle\left(\frac{1+z}{\delta}\right)^{4+2\alpha}\left(\frac{d_{L}}{\rm Mpc}\right)^{2}\left(\frac{\Delta T}{\rm h}\right)^{-1} \\
\left(\frac{F_{\rm 1\,keV}}{\mu \rm Jy}\right)\left(\frac{E_{\gamma}}{\rm GeV}\right)^{\alpha}
\end{aligned}
\label{eq4}
\end{equation} \citep{Fan13(a)}. 
Therefore, the lower limit on $\gamma$-ray Doppler factor can be estimated if we assume that the optical depth does not exceed unity,
\begin{equation}
\begin{aligned}
\delta_{\gamma}\ge\left[{1.54\times10^{-3}\displaystyle\left(1+z\right)^{4+2\alpha}\left(\frac{d_{L}}{\rm Mpc}\right)^{2}\left(\frac{\Delta T}{\rm h}\right)^{-1}} \right. \\
\left. \left(\frac{F_{\rm 1\,keV}}{\mu \rm Jy}\right){\left(\frac{E_{\gamma}}{\rm GeV}\right)^{\alpha}}\right]^{\frac{1}{4+2\alpha}}
\end{aligned}
\label{eq5}
\end{equation} \citep{Mat93, Fan13(a), Fan14}. 

The lower limit on $\gamma$-ray Doppler factor $\delta_{\gamma}$ can be calculated if the knowledge of the luminosity distance $d_{L}$ and redshift $z$, X-ray behaviour (characterized by the spectral index $\alpha_{X}$ and flux density $F_{\rm 1\,keV}$), $\gamma$-ray behaviour (characterized by the average $\gamma$-ray photon energy $E_{\gamma}$), and the timescale of variation $\Delta T$ are given.

\section{SAMPLE AND RESULTS} 

\subsection{Sample}

We compiled a catalog of 809 {\it Fermi}-detected blazars based on the identified by 4FGL with available X-ray data and present their derived lower limit on $\gamma$-ray Doppler factors in this work. 

For probing the origin of X-ray emission, \citet{Yan19} collected 660 $\gamma$-ray loud blazars from \citet{Fan16} with available X-ray data, which contained 269 FSRQs and 391 BL Lacs, to investigate the contributions from the synchrotron radiation and inverse Compton scattering to the X-ray emission in the $\gamma$-ray blazars, and obtained that they can be simply separated by their SED fitting curves from radio to X-ray bands by adopting a parabolic function, $\log(\nu F_{\nu})=P_{1}(\log\nu-\log\nu_{\rm p})^{2}+\nu_{\rm p}F_{\nu_{\rm p}}$, where $P_{1}$ is the spectral curvature, $\log\nu_{\rm p}$ and $\nu_{\rm p}F_{\nu_{\rm p}}$ denote the peak frequency and peak flux, respectively. Recently, \citet{Pei20(a)} compiled a large catalog of 4388 AGNs with available core-dominance parameters, $\log \, R$, in which 584 are {\it Fermi}-detected blazars based on the 4FGL.  

We adopt the X-ray data from \citet{Yan19} for 660 sources. For the rest 149 sources, we compiled their X-ray data via NED (NASA/IPAC Extragalactic Database\footnote{http://ned.ipac.caltech.edu/}), BZCAT (The Roma BZCAT-5th edition, Multi-frequency Catalogue of Blazars\footnote{http://www.asdc.asi.it/bzcat/}) \citep{Mas15}, and \citet{Fan14}. Finally we collected 809 $\gamma$-ray blazars, 342 are FSRQs and 467 are BL Lacs. According to the classification we described above \citep[see][]{Fan16, Bot19}, 467 BL Lacs are grouped into 202 HBLs (HSP BL Lacs), 213 IBLs (ISP BL Lacs) and 52 LBLs (LSP BL Lacs), respectively. We then cross-check these $\gamma$-ray blazars with \citet{Pei20(a)} and found 507 sources with available core-dominance parameter $\log R$, which includes 263 FSRQs and 244 BL Lacs. 

\subsection{Calculation}  

For a $\gamma$-ray source, the $K$-corrected $\gamma$-ray luminosity can be calculated from the detected photons \citep{Abd10b, Fan13(b)},
\begin{equation}
L_{\gamma} = 4\pi d_{\rm L}^{2}(1+z)^{\alpha_{\gamma}^{\rm ph}-2}f,
\label{eq6}
\end{equation} 
where $\alpha^{\rm ph}_{\gamma}$ is the $\gamma$-ray photon spectral index. The integral flux $f$ in units of GeV cm$^{-2}$ s$^{-1}$ can be obtained by $f=\int_{E_L}^{{E_U}}EdN$, and we adopt $E_{L}=1$ GeV and $E_{U}=100$ GeV respectively in our calculation.  
   
For the sources whose X-ray spectral index is not given, we took the median of $\alpha_{X}$ for subclasses into account, 1.022 for FSRQs and 1.008 for BL Lacs. If the redshift is not available, we then use the average values of the subsample to substitute it, i.e. $\langle z\rangle|_{\rm FSRQ}=1.172$ and $\langle z\rangle|_{\rm BL\;Lac}=0.499$. The average $\gamma$-ray photon energy $E_{\gamma}$ can be calculated by $\langle E\rangle=\int EdN/\int dN$. The variability timescales for most sources are unknown, even though a few ones are available \citep{YF10}. In our calculation, for the sake of simplicity, we adopt $\Delta T$=1 day \citep{DG95, Ghi98, Fan13(a), Fan14}. Consequently, we can calculate the lower limit on $\gamma$-ray Doppler factor $\delta_{\gamma}$.  

\subsection{Results} 

From our calculations, we obtained the average value of $\gamma$-ray Doppler factors for our whole sample, $\langle\delta_{\gamma}\rangle|_{\rm blazar}=5.39\pm3.70$. For 342 FSRQs, we ascertain that their $\gamma$-ray Doppler factor, on average, is $\langle\delta_{\gamma}\rangle|_{\rm FSRQ}=6.87\pm4.07$, ranging from $\delta_{\gamma}=1.04$ of J0625.8-5441 to $\delta_{\gamma}=28.38$ of J1833.6-2103. On the other hand, BL Lacs have the $\gamma$-ray Doppler factor, on average, $\langle\delta_{\gamma}|_{\rm BL\,Lac}\rangle=4.31\pm2.97$ in the range from $\delta_{\gamma}=0.95$ of J0113.7+0225 to $\delta_{\gamma}=22.81$ of J2055.4-0020. We present our sample and results in Table \ref{tab1}. In this table, Col. 1 gives 4FGL name; Col. 2 other name; Col. 3 classification (FSRQ: flat spectrum radio quasar; HBL: high synchrotron peak BL Lacs; IBL: intermediate synchrotron peak BL Lacs; LBL: low synchrotron peak BL Lacs); Col. 4 redshift; Col. 5 core-dominance parameter; Col. 6 the X-ray flux density in units of $\mu$Jy at 1 keV; Col. 7 X-ray spectral index; Col. 8 Reference for Col. 6 and 7; Col. 9 $\gamma$-ray photon index; Col. 10 average $\gamma$-ray photon energy in units of GeV; Col. 11 X-ray luminosity in units of erg s$^{-1}$; Col. 12 $\gamma$-ray luminosity in units of erg s$^{-1}$; Col. 13 the derived lower limit on $\gamma$-ray Doppler factor in this paper; Col. 14 the estimated Doppler factor from \citet{Lio18}; Col. 15 the estimated Doppler factor from \citet{Che18}. This table is available in its entirety in machine-readable form.

\begin{table*}
\caption{The lower limit on $\gamma$-ray Doppler factor for {\it Fermi} blazars} 
\centering
\begin{tabular*}{\textwidth}{@{}c\x c\x c\x c\x c\x c\x c\x c\x c\x c\x c\x c\x c\x c\x c@{}}
\hline \hline
4FGL Name & Other Name & Class & z & $\log R$ & $F_{\rm 1\,KeV}$ & $\alpha_{X}$ & Ref. & $\alpha^{\rm ph}_{\gamma}$ & $E_{\gamma}$ & $L_{X}$ & $L_{\gamma}$ & $\delta_{\gamma}$ & $\delta_{\rm L18}$ & $\delta_{\rm C18}$ \\
(1) & (2) & (3) & (4) & (5) & (6) & (7) & (8) & (9) & (10) & (11) & (12) & (13) & (14) & (15)\\
\hline
J0005.9+3824 	&	0003+380	&	FSRQ	&	0.229	&	1.13 	&	0.080 	&	1.32 	&	Y19	&	2.67 	&	2.39 	&	43.44 	&	44.46 	&	1.92 	&	5.23	 &  5.6 \\
J0006.3-0620 	&	0003-066 	&	HBL	&	0.347 	&	0.26 	&	0.152 	&		&	NED	&	2.17 	&	3.75 	&	44.14 	&	44.48 	&	2.95 	&	6.96 &	\\
J0010.6+2043 	&	0007+205	&	FSRQ	&	0.6	&	0.29 	&	0.058 	&		&	NED	&	2.32 	&	3.20 	&	44.30 	&	45.12 	&	3.56 	&	6.02	 & \\
J0019.6+7327 	&	0016+731	&	FSRQ	&	1.781	&	0.54 	&	0.015 	&		&	NED	&	2.59 	&	2.51 	&	44.88 	&	47.30 	&	6.96 	&	7.84 & 	\\
J0050.7-0929 	&	0048-097	&	IBL	&	0.634	&	1.20 	&	0.392 	&	1.57 	&	Y19	&	2.04 	&	4.41 	&	45.19 	&	46.64 	&	5.12 	&	20.23 &	28.4 \\
J0108.6+0134 	&	0106+013	&	FSRQ	&	2.099	&	0.71 	&	0.065 	&	0.43 	&	Y19	&	2.35 	&	3.08 	&	45.69 	&	48.37 	&	14.45 	&	2.64 &	15.3 \\
J0113.4+4948 	&	0110+495	&	FSRQ	&	0.389	&	0.98 	&	0.104 	&	2.24 	&	Y19	&	2.23 	&	3.50 	&	44.10 	&	45.49 	&	2.86 	&	5.66 & 	10.6 \\
J0116.0-1136 	&	0113-118	&	FSRQ	&	0.670 	&	1.02 	&	0.180 	&	0.98 	&	Y19	&	2.39 	&	2.97 	&	44.91 	&	46.23 	&	5.18 	&	8.22	 & 9.9 \\
J0132.7-1654 	&	0130-171	&	FSRQ	&	1.020 	&	0.36 	&	0.034 	&		&	Y19	&	2.40 	&	2.96 	&	44.64 	&	46.53 	&	4.55 	&	12.53 &	18 \\
J0137.0+4751 	&	0133+476	&	FSRQ	&	0.859	&	0.91 	&	0.324 	&	0.82 	&	Y19	&	2.29 	&	3.29 	&	45.43 	&	46.98 	&	6.85 	&	12.73  &	31.6 \\
J0141.4-0928 	&	0138-097	&	IBL	&	0.733 	&	-0.04 	&	0.045 	&	1.15 	&	Y19	&	2.17 	&	3.77 	&	44.40 	&	46.45 	&	4.17 	&	15.19 &	14.3 \\
J0152.2+2206 	&	0149+218	&	FSRQ	&	1.320 	&	0.69 	&	0.048 	&		&	BZCAT	&	2.71 	&	2.32 	&	44.54 	&	45.88 	&	5.06 	&	4.32 &	10.2\\
$\cdots$ & $\cdots$ & $\cdots$ & $\cdots$ & $\cdots$ & $\cdots$ & $\cdots$ & $\cdots$ & $\cdots$ & $\cdots$ & $\cdots$ & $\cdots$ & $\cdots$ & $\cdots$ & $\cdots$\\
\hline \hline
\end{tabular*}\label{tab1}

\medskip
\tabnote{Note: Col. 1 gives 4FGL name; Col. 2 counterpart name; Col. 3 classification (FSRQ: flat spectrum radio quasar; HBL: high synchrotron peak BL Lacs; IBL: intermediate synchrotron peak BL Lacs; LBL: low synchrotron peak BL Lacs); Col. 4 redshift; Col. 5 core-dominance parameter; Col. 6 the X-ray flux density in units of $\mu$Jy at 1 keV; Col. 7 X-ray spectral index; Col. 8 Reference for Col. 6 and 7 (Y19: \citet{Yan19}; NED: NASA/IPAC Extragalactic Database; BZCAT: The Roma BZCAT- 5th edition, Multi-frequency Catalogue of Blazars); Col. 9 $\gamma$-ray photon index; Col. 10 average $\gamma$-ray photon energy in units of GeV; Col. 11 X-ray luminosity in units of erg s$^{-1}$; Col. 12 $\gamma$-ray luminosity in units of erg s$^{-1}$; Col. 13 the derived lower limit on $\gamma$-ray Doppler factor; Col. 14 the estimated Doppler factor from \citet{Lio18}; Col. 15 the estimated Doppler factor from \citet{Che18}.    

(The table is available in its entirety in machine-readable form. Readers can also send your request to {\it zhiyuan.pei@phd.unipd.it})}
\end{table*}

\begin{figure}
   \centering
   \includegraphics[width=9cm]{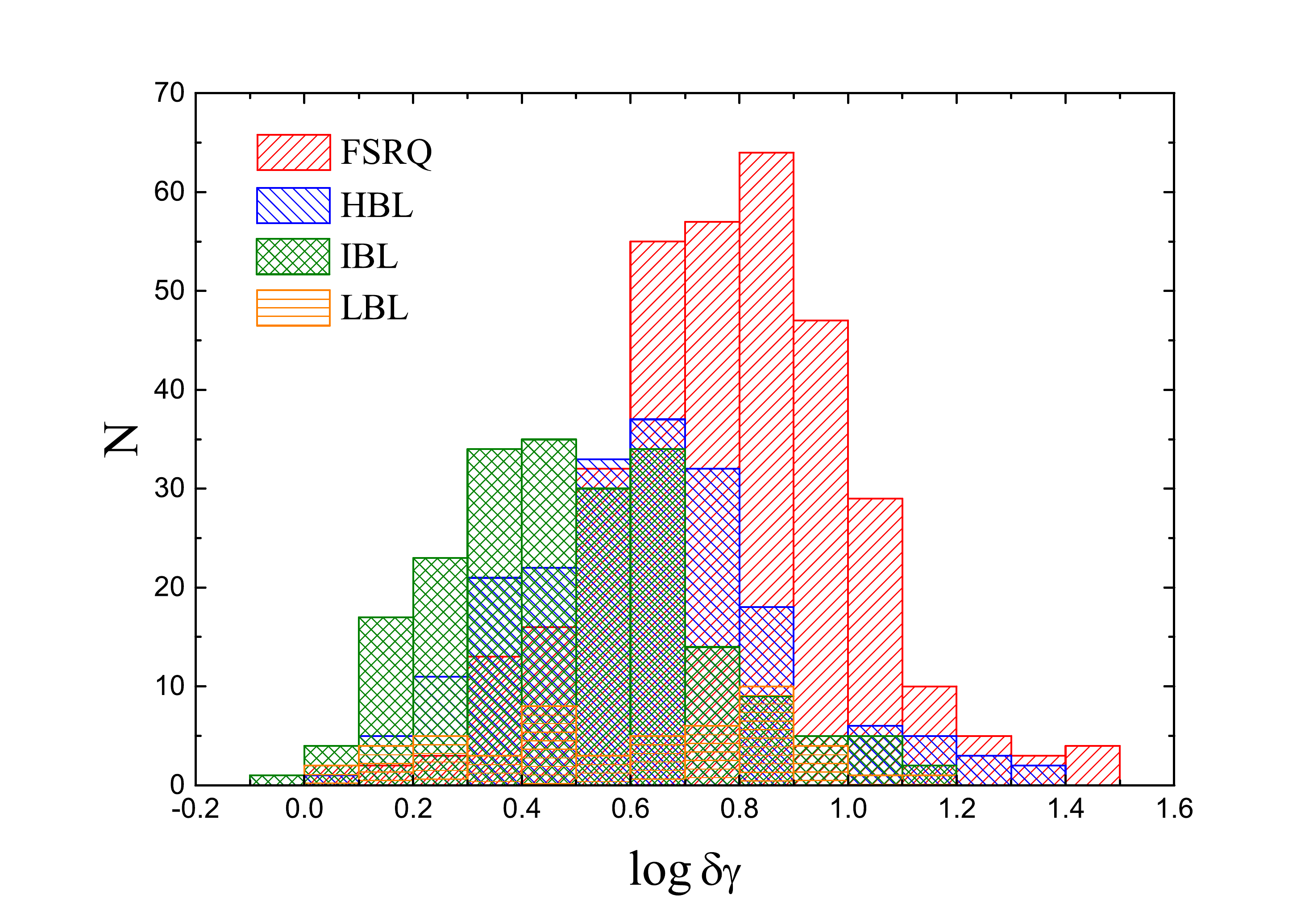}
   \caption{Distributions of the lower limit on $\gamma$-ray Doppler factor ($\delta_{\gamma}$) in logarithm for all subclasses.}
     \label{fig1}
\end{figure}

The distributions of $\delta_{\gamma}$ in logarithm for FSRQs and different classes of BL Lacs are shown in Figure \ref{fig1}. A Kolmogorov-Smirnov test (hereafter, K-S test) between the distributions of $\delta_{\gamma}$ for FSRQs and BL Lacs shows that they belong to different parent distributions ($p=2.54\times10^{-27}$). From the distributions and the K-S test result, we can find that $\langle \delta_{\gamma} \rangle|_{\rm FSRQ}> \langle \delta_{\gamma} \rangle|_{\mathrm{BL\,Lac}}$, indicating that the {\it Fermi}-detected FSRQs are more $\gamma$-ray Doppler boosted.     

\section{DISCUSSION}

Blazars, as the subclass of AGNs, show extreme observational properties, which are associated with the relativistic beaming effect. All of these extreme properties indicate that blazars are the most active extragalactic sources in the Universe. BL Lac objects are usually identified as ``lineless'' active galactic nuclei and conversely, quasars show strong broad emission lines. The core-dominance parameter, $R$, can be used for the orientation indicator of the jet \citep{UP95}. Since the Doppler factor is not observable and cannot be determined accurately, thus the core-dominance parameter might be an eligible indicator of Doppler beaming effect.    

As ever, blazars take a majority of sources detected by {\it Fermi} \citep{Abd09, Abd10a, Abd10b, Ace15}. \citet{Pei20(a)} present a up-to-date largest catalog of available core-dominance parameters $R$. We point out that $R$ are quite different for diverse subclasses of AGNs, and particularly, {\it Fermi} blazars hold, on average, higher $R$ than the non-{\it Fermi} blazars, indicating that the $\gamma$-emissions of {\it Fermi} blazars are from the jet and more Doppler boosted. The $\gamma$-emission is strongly beamed \citep{Fan09, Pei16, Pei20(a)}.  

The strongly Doppler boosted emission is referable to the relativistic beaming effect which enhances the observed flux density by a factor of $\delta^{2+\alpha}$ for stationery and continuous jet, and $\delta^{3+\alpha}$ for a moving blob. Since the $\gamma$-ray blazars are transparent to $\gamma\gamma$ pair production within a small region deducted from the fast $\gamma$-ray variability, strongly suggesting that the $\gamma$-ray emission produced from the jets of blazars is also Doppler beamed, which is similar to the behaviour of radio emission \citep{Mon95, Mat93, Fan09}. Therefore, the estimation of $\gamma$-ray Doppler factor is reasonable and substantial to explore the typical characteristic of $\gamma$-ray loud blazars and beaming effect \citep{Fan13(a), Fan14}.         

\subsection{Comparison with other Doppler factors in the literature}   

The Doppler factor ($\delta$), an important parameter to reveal the relativistic beaming effect and explain the observed extreme properties of blazars, is proverbially unmanageable to measure since there is no straight method at present. Many indirect methods are proposed to estimate the beaming Doppler factor: (i) It can be deduced by a synchrotron self-Compton (SSC) model thus denoted $\delta_{\rm ssc}$ \citep[e.g.][]{Ghi93}; (ii) To be derived from adopting single-epoch radio data by assuming that the sources hold an equipartition of energy between radiating particles and magnetic field as denoted $\delta_{\rm eq}$ \citep{Rea94}; (iii) To be estimated using the radio flux density variations or brightness temperature denoted $\delta_{\rm var}$ \citep{LV99, Hov09}. (iv) One can calculate it based on the broadband SED \citep[e.g.][]{Che18}. However, due to these different assumptions, each method will render discrepant results. 

The relativistic beaming effect plays a crucial role in the $\gamma$-ray emission, and particularly, for {\it Fermi}-detected blazars. \citet{Fan09} found that the $\gamma$-ray luminosity of {\it Fermi}-detected blazars correlates tightly with the radio Doppler factor $\log L_{\gamma} \sim 0.47\log \delta^{3+\alpha}_{\rm R}$. \citet{Kov09} pointed out that the sources detected by {\it Fermi}/LAT have higher brightness temperature with respect to those not detected by {\it Fermi}. \citet{Sav10} compiled 62 AGNs with apparent superluminal motion and adopted their Doppler factors from \citet{Hov09}, and found that the {\it Fermi} blazars have, on average, higher Doppler factor than non-{\it Fermi}-detected blazars.
\citet{Xia19} collected 291 sources with superluminal motions, in which 189 are $\gamma$-ray sources detected by {\it Fermi}, and reported that the {\it Fermi}-detected sources show higher proper motion, apparent velocity, Doppler factor, Lorentz factor and smaller viewing angles than non-{\it Fermi}-detected sources, also suggesting the strong Doppler effect lies on those $\gamma$-ray sources. \citet{Pei20(a)} obtained that the $\gamma$-ray luminosity increases with radio core-dominance parameter for {\it Fermi} AGNs.

Beaming effect is mostly studied using the radio emission, which yields that the radio variability Doppler factor ($\delta_{\rm var}$) method should be perhaps an appropriate way to describe the blazars population and beaming effect \citep{Fan09, Fan13(b), Lio15, Lio17(a), Lio17(b)}.  

\begin{figure*}
   \centering
   \includegraphics[width=8.6cm]{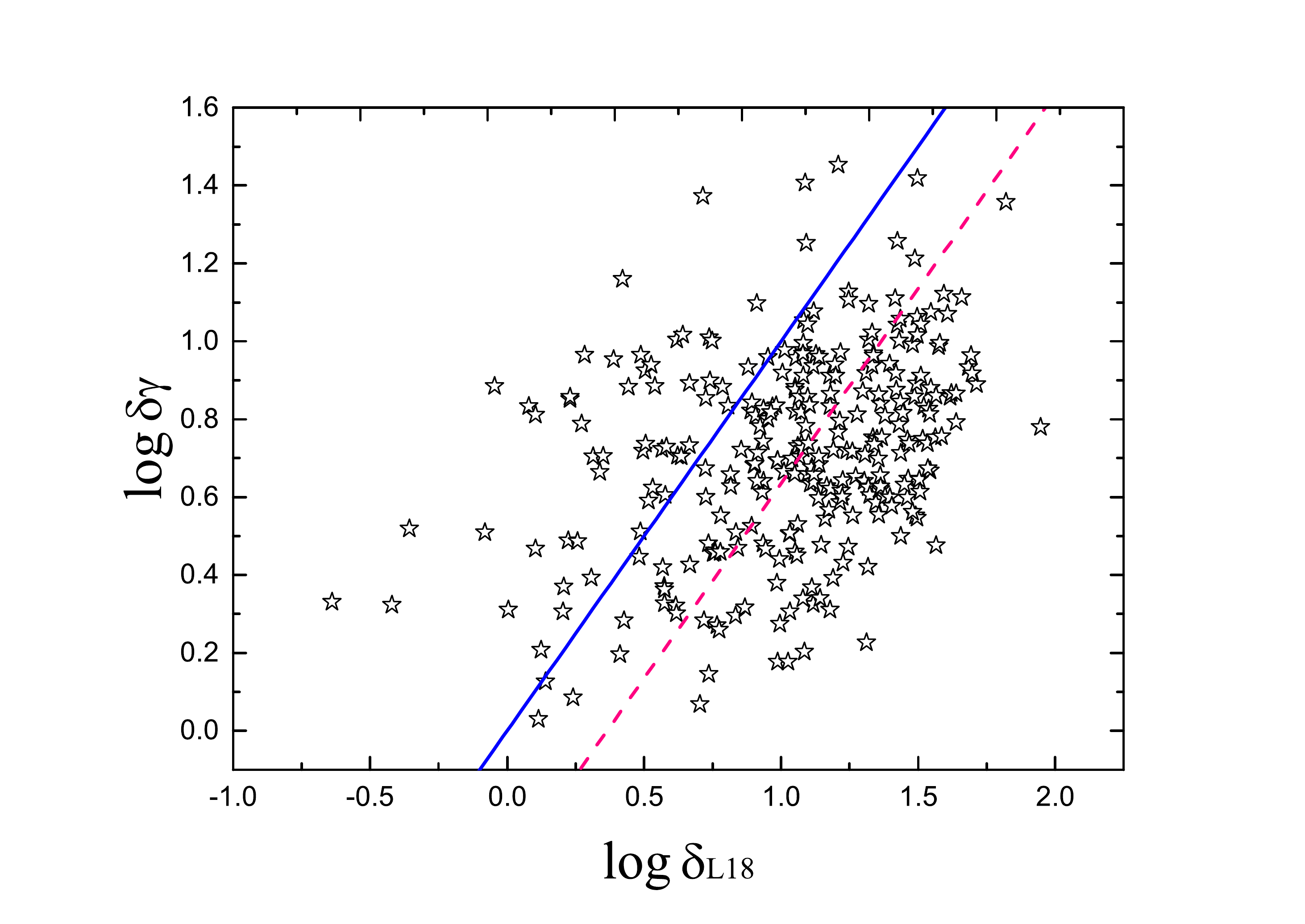}
   \includegraphics[width=8.7cm]{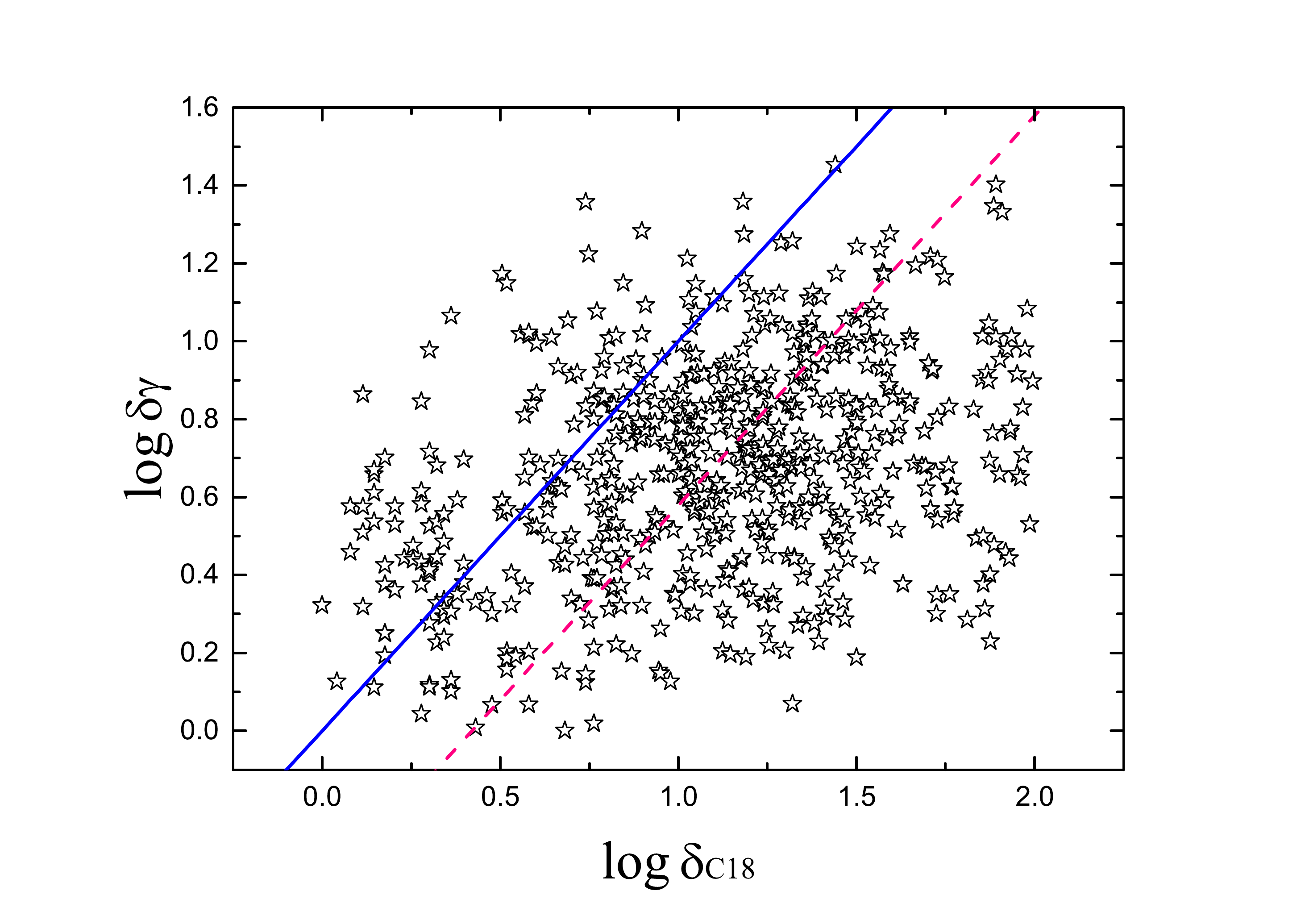}
   \caption{Plot of the correlation between $\log \delta_{\gamma}$ derived in this paper and that presented from other literature after cross-checking. $\log \delta_{\rm L18}$ denotes the variability Doppler factor adopted from \citet{Lio18} (left panel) and $\log \delta_{\rm C18}$ denotes the SED fitting derived Doppler factor from \citet{Che18} (right panel). The solid blue lines refer to the equality line and the dashed pink ones signify the half proportion dividing line that are parallel to the equality one.} 
     \label{fig2}
\end{figure*}

By modeling the radio light curves of 1029 sources as a series of flares characterized by an exponential rise and decay, \citet{Lio18} estimated the variability Doppler factor ($\delta_{\rm var}$) for 837 blazars, which included 670 FSRQs and 167 BL Lacs. They calculated the variability brightness temperature ($T_{\rm var}$) using
\begin{equation}
T_{\rm var} = 1.47\times10^{13}\displaystyle\frac{d^{2}_{\rm L}\Delta S_{\rm ob}(\nu)}{\nu^{2}t^{2}_{\rm var}(1+z)^{4}}{\rm K},
\label{eq7}
\end{equation}     
here, $S_{\rm ob}(\nu)$ the amplitude of the flare in Jy, $\nu$ the observed frequency in GHz and $t_{\rm var}$ the rise time of a flare in days. Then the variability Doppler factor ($\delta_{\rm var}$) can be defined as
\begin{equation}
\delta_{\rm var}=(1+z)\sqrt[3]{\frac{T_{\rm var}}{T_{\rm eq}}},
\label{eq8}
\end{equation}     
where $T_{\rm eq}$ is the equipartition brightness temperature, and $T_{\rm eq}=2.78\times10^{11}$K was adopted. After cross-checking with our sample, there are 285 common sources, which contains 210 FSRQs and 75 BL Lacs. When we compared our results with theirs, it was found that $\log \delta_{\gamma}=(0.22\pm0.03)\log \delta_{\rm L18}+(0.49\pm0.04)$ with a correlation coefficient $r=0.37$ and a chance probability of $P<10^{-10}$. We show this plot in the left panel of Figure \ref{fig2}. 

Based on the SED-fitting, \citet{Che18} estimated the jet physical parameters of 1392 $\gamma$-ray loud blazars taking from \citet{Fan16}, and particularly, they calculated the Doppler factor ($\delta_{\rm SED}$) and obtained that the median values of the Doppler factors of FSRQs, BL Lacs, and total blazars were 10.7, 22.3, and 13.1, respectively. It is usually assumed in SED modeling that the $\gamma$-ray emission is produced closer to the supermassive black hole than the radio core of the jet where most of the radio emission originates. For comparison, we investigate the relation between the derived $\delta_{\gamma}$ in the present paper and those in \citet{Che18} for 597 sources are in common in both papers (In fact, we   found 682 sources to be in common after cross-checking, however, there are 85 sources obtained an extremely large or small value of $\delta_{\rm SED}$ in \citet{Che18}, thus we excluded those 85 sources). The best-fitting is $\log \delta_{\gamma}=(0.21\pm0.02)\log \delta_{\rm C18}+(0.45\pm0.03)$ with $r=0.34$ and $P\sim0$ for the 597 sources (see the right panel of Figure \ref{fig2}).

However, according to equation (\ref{eq5}), the estimation of lower limit on $\delta_{\gamma}$ would be affected by redshift, the above correlations then have a redshift dependence, thus the redshift effect needs to be removed. To do so, we adopt a partial correlation analysis \citep[see e.g.][]{Pad92}, 
\begin{equation}
r_{12,3} = \displaystyle\frac{r_{12}-r_{13}r_{23}}{\sqrt{1-r^{2}_{13}}\sqrt{1-r^{2}_{23}}}
\label{eq9}
\end{equation} 
where $r_{ij}$ denotes the correlation coefficient between $x_{i}$ and $x_{j}$, whilst $r_{ij,k}$ denotes the partial correlation coefficient between $x_{i}$ and $x_{j}$ with $x_{k}$ dependence excluded ($i, j, k=1,2,3$). In our case, we let $x_{1}=\log\delta_{\gamma}$, $x_{2}=\log\delta_{\rm L18}$ or $\log\delta_{\rm C18}$  and $x_{3}=z$. For the left panel, we have $r_{12}=0.37$, $r_{1z}=0.74$ and $r_{2z}=0.31$, which yields $r_{12,z}=0.21$. Using the similar calculation, we obtain $r_{12,z}=0.19$ for the right panel. Their $P$-values are both $<10^{-4}$. It still rendered a statistically correlation between our derived values of $\log\delta_{\gamma}$ and those from other methods after removing the redshift effect, implying that they are truly correlated. This result indicates that our derived lower limit on $\gamma$-ray Doppler factors are reasonable.   

We draw an equality line in both panels of Figure \ref{fig2} (labeled in blue solid), and note that some points are quite disperse and most importantly, the derived values of $\log \delta_{\gamma}$ are fairly small than that of $\log \delta_{\rm var}$ or $\log \delta_{\rm SED}$ and the range is relatively compact, ranging from 0.95 to 28.38 and the average value is 5.39. We should point out that, firstly, because our estimation is the lower limit on $\delta_{\gamma}$. \citet{Hov09} had shown the median value of $\delta$ was 12.02 ranging from 0.30 to 35.50; \citet{Lio18} reported the average value to be 14.35 in the range from 0.08 to 88.44; \citet{Che18} on average, 14.30 was obtained and spanning from 1.00 to 99.50. The typical values of Doppler factor in blazars should be in the range from a few to $\sim$50 based on various methods as mentioned above. However, since our derived results are the lower limit on $\delta_{\gamma}$, thus the distribution of $\delta_{\gamma}$ should be smaller than those from the literature. This is consistent with \citet{DG95}, they found the average value $\langle\delta_{\gamma}\rangle\sim4.4$, ranging from 1.3 to 11 for a sample of EGRET blazars. \citet{Fan13(a)} obtained an average value $\langle\delta_{\gamma}\rangle\sim7.22$ for 138 $\gamma$-ray blazars and \citet{Fan14} also found the average value was $\sim$7.00 with regard to their sample. 

Secondly, since we could not ascertain the variability timescale for each source in our large sample, thus we adopted $\Delta T=24$ hs (1 day) in our present calculation. However, this operation was examined by \citet{Fan13(a)} and \citet{Fan14}, who also used $\Delta T=1$ day for calculation and obtained reliable results. For instance, those authors found a tendency for the $\gamma$-ray Doppler factors to increase with the radio Doppler factors and $\delta_{\gamma}$ are also correlated with the superluminal velocity. This supports the fact that the $\gamma$-rays are strongly beamed, and likewise suggesting that the radio Doppler factors estimated from the variability can be used to discuss the beaming effect in {\it Fermi} loud blazars. \citet{DG95} used a similar method as this paper to estimate the lower limit on $\delta_{\gamma}$ for 46 $\gamma$-ray loud blazars. They could not find enough information of variability timescale for nearly half of their sample either, and they also adopted $\Delta T=1$ day for calculation.  However, different synchrotron peaked sources may have different timescales. \citet{LAT11}, \citet{Bon11}, \citet{Hu14} and \citet{Pri20} processed systematic well-studied on Mrk 421, 3C 454.3, S5 0716+714 and 3C279, respectively, although Mrk 421 is a high synchrotron peak blazar and the other three are low synchrotron peak blazars \citep{Fan16}, those authors had shown that a typical variability timescale in the source frame for {\it Fermi}/LAT blazars is $\approx$1 day \citep[see also][]{Nal13, Zha15}. \citet{Che18} also adopted this idea for simplicity in the constraint of Doppler factor from SED modeling for a larger sample of blazars \citep[see also][]{Kan14}. From this point of view, we consider that our choice of variability timescale that $\Delta T=1$ day is also reliable.  

On the other hand, therefore, different values of $\Delta T$ can be also considered. We compute the $\gamma$-ray Doppler factor with $\Delta T=6$ hs (denoted in $\delta_{\gamma}^{\rm6\,hs}$) and $\Delta T=48$ hs (denoted in $\delta_{\gamma}^{\rm48\,hs}$), then a relation is found $\delta_{\gamma}^{\rm6\,hs}\sim1.32\delta_{\gamma}^{\rm24\,hs}$ and $\delta_{\gamma}^{\rm48\,hs}\sim0.87\delta_{\gamma}^{\rm24\,hs}$. A slight higher than our present result if we choose $\Delta T=6$ hs and smaller while $\Delta T=48$ hs is considered \citep[see also][]{Fan14}. For the sake of simplicity, we adopt $\Delta T=24$ hs in this paper for calculation as did in \citet{DG95}, \citet{Fan13(a)} and \citet{Fan14}. Besides, the non-simultaneous observations will also result in some discrepancies. 

For comparison, we draw a half proportion dividing line that is parallel to the equality line in two panel (labeled in dashed pink), which means there are 50/50 distribution of sources on two sides. When we consider this line in the plot of $\log \delta_{\gamma}$ versus $\log \delta_{\rm L18}$, it corresponds to a variability timescale $\Delta T \sim$ 2 hs. For the case of $\log \delta_{\gamma}$ versus $\log \delta_{\rm C18}$, we found $\Delta T \sim$ 1.5 hs. This could suggest that the on average variability timescale for {\it Fermi}-detected blazars is around 1.5$\sim$2 hs, however, we cannot reach firm conclusions.         

Based on the synchrotron self-Compton limit, a so-called ``classical'' method to estimate the Doppler factor, \citet{Ghi93} constrained a relation between the $\delta$ derived from moving sphere and a continuous jet,
\begin{equation}
\delta_{3+\alpha}^{(4+2\alpha)/(3+2\alpha)}=\delta_{2+\alpha}.
\label{eq10}
\end{equation} 

We also assume that the emission region from the jet is spherical, which means $p=3+\alpha$ has been taken into account. Note that, if the source is a continuous jet, $p=2+\alpha$, then the equation (\ref{eq5}) of derived $\delta_{\gamma}$ should be regenerated as     
\begin{equation}
\begin{aligned}
\delta_{\gamma}\ge\left[{1.54\times10^{-3}\displaystyle\left(1+z\right)^{3+2\alpha}\left(\frac{d_{L}}{\rm Mpc}\right)^{2}\left(\frac{\Delta T}{\rm h}\right)^{-1}} \right. \\
\left. \left(\frac{F_{\rm 1\,keV}}{\mu \rm Jy}\right){\left(\frac{E_{\gamma}}{\rm GeV}\right)^{\alpha}}\right]^{\frac{1}{3+2\alpha}}.
\end{aligned}
\label{eq11}
\end{equation} 
For comparison, we also estimate the lower limit on $\gamma$-ray Doppler factor for this case and obtain $\langle\delta_{\gamma}|_{\rm FSRQ}\rangle=8.93\pm6.78$ and $\langle\delta_{\gamma}|_{\rm BL\,Lac}\rangle=5.42\pm4.29$ with $P$-value of K-S test indicating the two distributions being from the same apparent distribution is $P=4.58\times10^{-35}$. Our results show that $\log\delta_{\gamma}^{2+\alpha}\sim1.14\log\delta_{\gamma}^{3+\alpha}$ for FSRQs and $\log\delta_{\gamma}^{2+\alpha}\sim1.13\log\delta_{\gamma}^{3+\alpha}$ for BL Lacs. We can find the estimated value of $\delta_{\gamma}$ for the case of continuous jet to be larger than that of moving blob sphere, and FSRQs also have on average higher $\gamma$-ray Doppler factor than BL Lacs.    

From our calculation, FSRQs have on average significantly higher $\delta_{\gamma}$ than BL Lacs ($P$-value is $2.54\times10^{-27}$ from K-S test), which is consistent with estimations from other literatures \citep{Fan09, Hov09, Fan13(a), Fan14, Lio18}. In the standard model of AGNs, FSRQs should have a smaller viewing angle $\theta$ with regard to BL Lacs, thus showing stronger Doppler boosted effect and resulting in a larger $\delta$ due to the definition of Doppler factor $\delta=[\Gamma(1-\beta\cos\theta)]^{-1}$. In the cross-checked sample with \citet{Lio18}, 157 out of 285 sources were given the estimation of viewing angle, which included 127 FSRQs with average around $\langle\theta\rangle|_{\rm FSRQ}\sim4.61$ deg and 30 BL Lacs with $\langle\theta\rangle|_{\rm BL\,Lac}\sim10.20$ deg, leading FSRQs have higher $\delta$ than BL Lacs.     

As for the weaker emission line features in BL Lacs is mainly ascribable to the fact that the isotropic emission component in BL Lacs is intrinsically weaker than in FSRQs and directs to different parent populations for the two subclasses of blazars \citep{MGP87, Ghi93}. For example, \citet{Pad92} pointed out that [O $\uppercase\expandafter{\romannumeral3}$] line luminosity and radio extended luminosity for BL Lacs are lower than for FSRQs with two orders of magnitude. It is possible that the different emission line property is from the factor that the ratio of the core emission to the extend one in the co-moving frame is higher in BL Lacs than in FSRQs \citep{Fan03}.

\subsection{Correlation analysis} 

\begin{figure*}
   \centering
   \includegraphics[width=8.6cm]{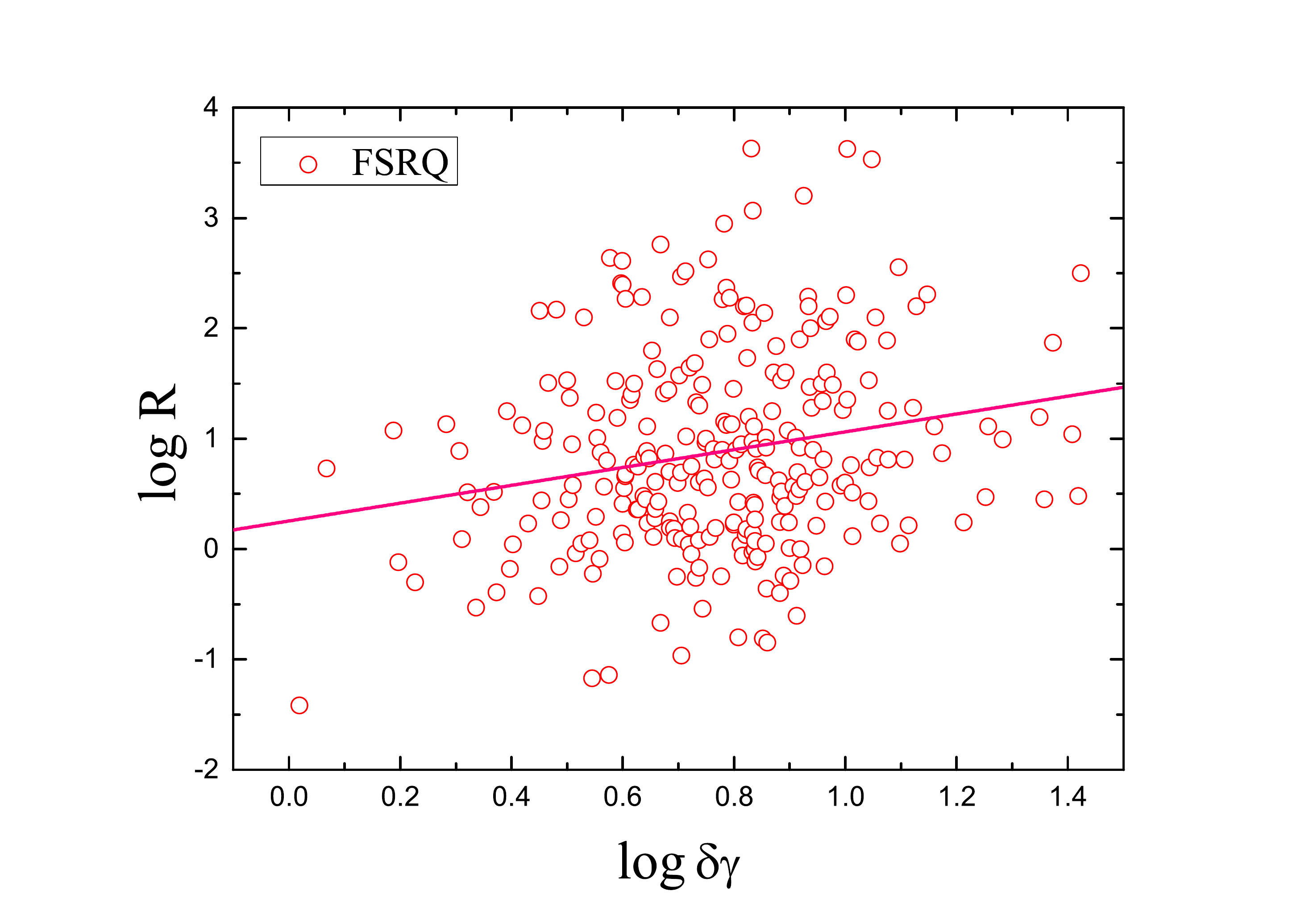}
   \includegraphics[width=8.6cm]{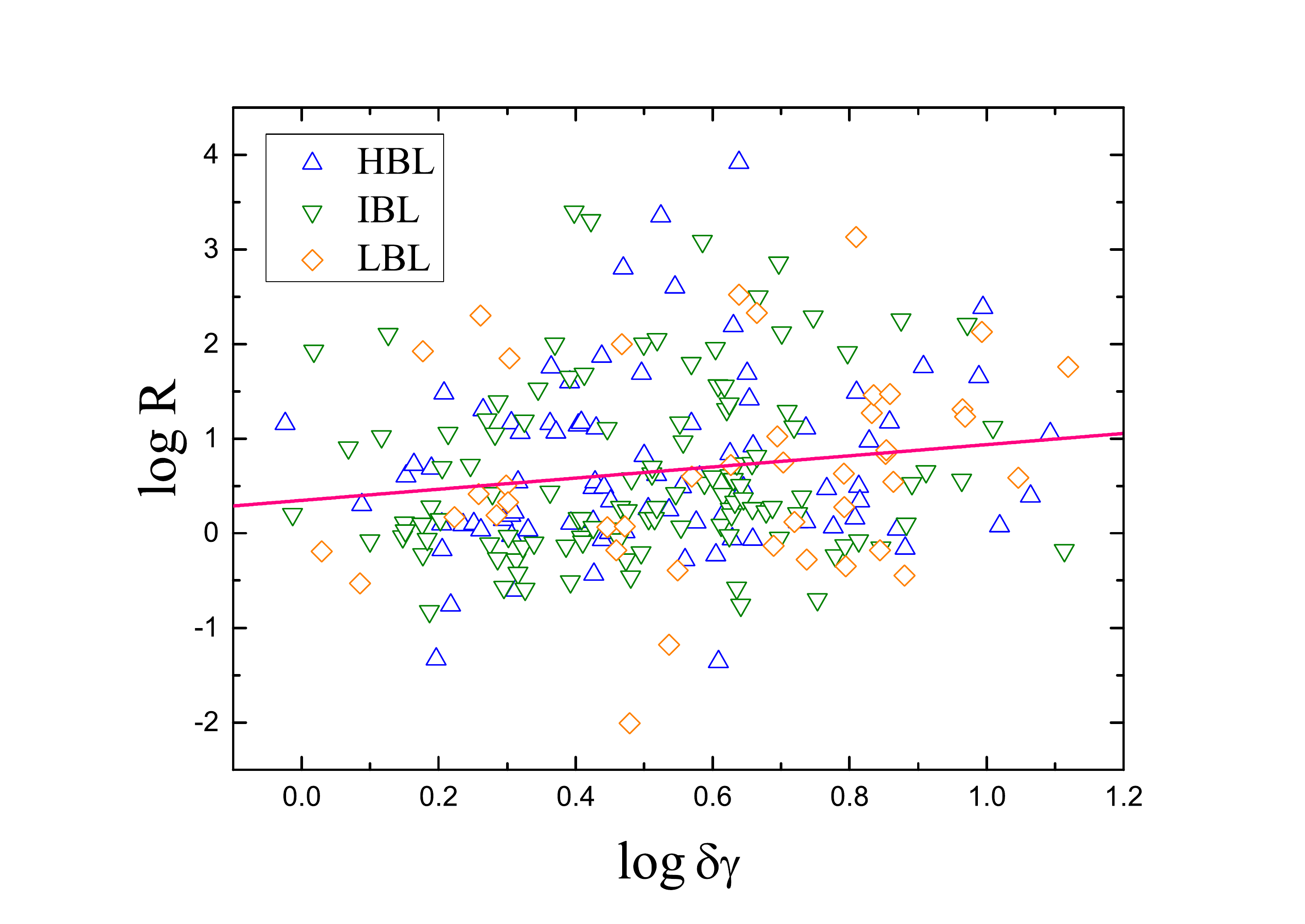}
   \caption{Plot of the core-dominance parameter $\log \, R$ against the  $\gamma$-ray Doppler factor $\log \, \delta_{\gamma}$ for FSRQs (left panel), and BL Lacs (right panel).}
     \label{fig3}
\end{figure*}

The beaming model anticipates that more core-dominated sources should be more beamed and consequently have larger Doppler boosted factors.
\citet{DG95} had probed the correlation between the core-dominance parameter $\log \, R$ and the $\gamma$-ray Doppler factor $\log \, \delta_{\gamma}$, a correlation coefficient $r=0.38$ and a chance probability of $P=8.0\times10^{-2}$ were found for 28 sources after excluding BL Lacs since their core-dominance parameters were quite large. The relation $R-\delta_{\gamma}$ existed but their sample is small. We also explore the correlation between $\log \, R$ and the derived $\log \, \delta_{\gamma}$ as the scatter plot in Figure \ref{fig3}. The best-fitting for FSRQs is 
$$\log R=(0.81\pm0.24)\log\delta_{\gamma}+(0.25\pm0.19)$$
with a correlation coefficient $r=0.22$ and a chance probability of $P<10^{-4}$;
$$\log R=(0.59\pm0.24)\log\delta_{\gamma}+(0.35\pm0.14)$$
with $r=0.16$ and $P=0.01$ for BL Lacs. The $R-\delta_{\gamma}$ relation shows that a $\gamma$-ray source with more core-dominated to be more Doppler beamed.   

\citet{Hov09} collected 80 sources with valid core-dominance parameters, to investigate the correlation between $\log R$ and Doppler factor they derived, and found a positive correlation with $r=0.37$ and $P=4.0\times10^{-4}$. \citet{Pei20(a)} studied $R-\delta$ relation by cross-checking the Doppler factor from \citet{Lio18} and a significant correlation was also obtained. This demonstrates that the indication of sources with more core-dominated are also more boosted. \citet{Ghi93} expressed the core-dominance parameter as $R = f\delta^{3+\alpha}$ for the sources with the jets directly pointing to the observer. Therefore, the core-dominance parameter is statistically a good indicator for beaming effect \citep{UP95, Fan03, Pei16, Pei19}.

\begin{figure*}
   \centering
   \includegraphics[width=8.6cm]{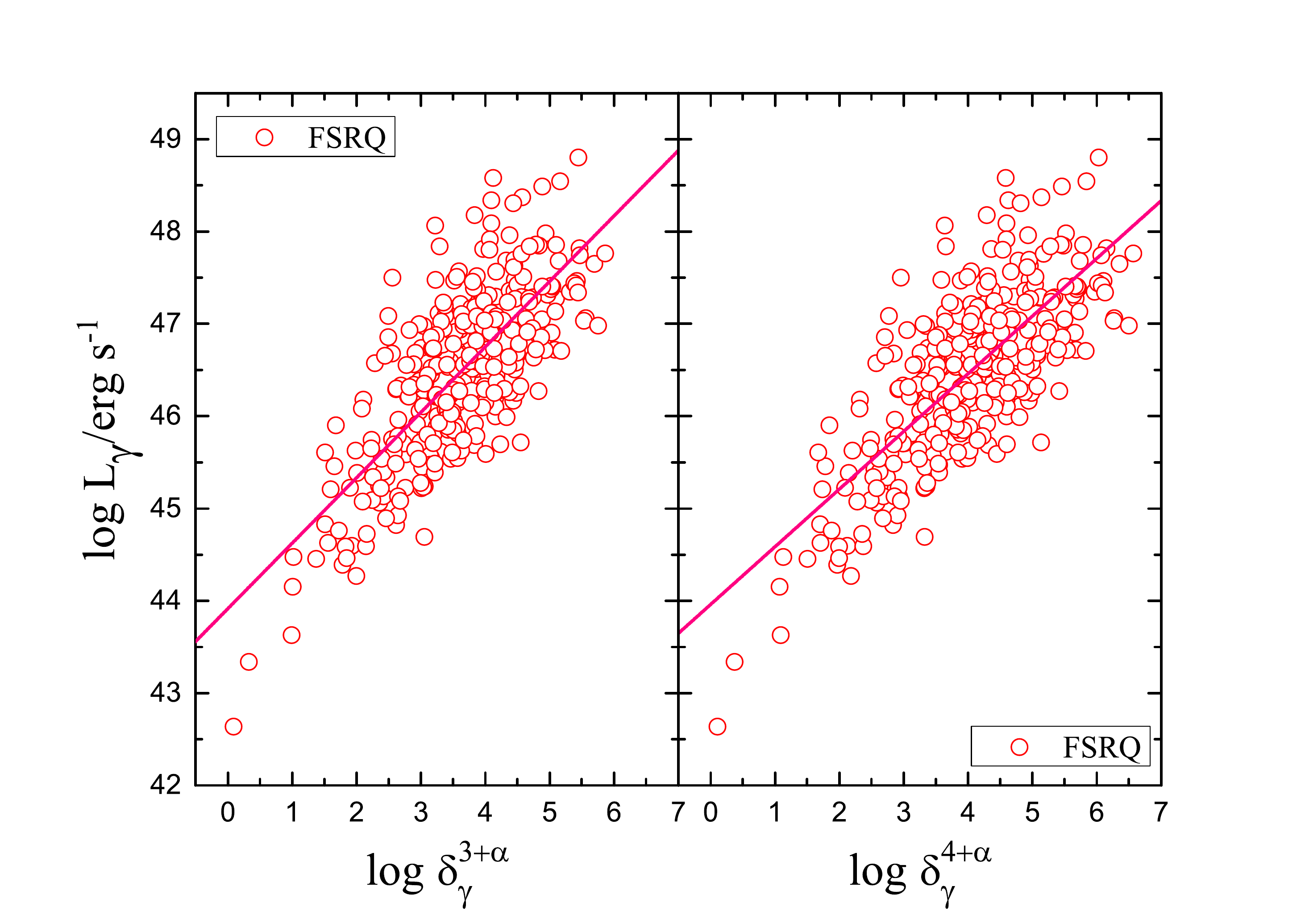}
   \includegraphics[width=8.6cm]{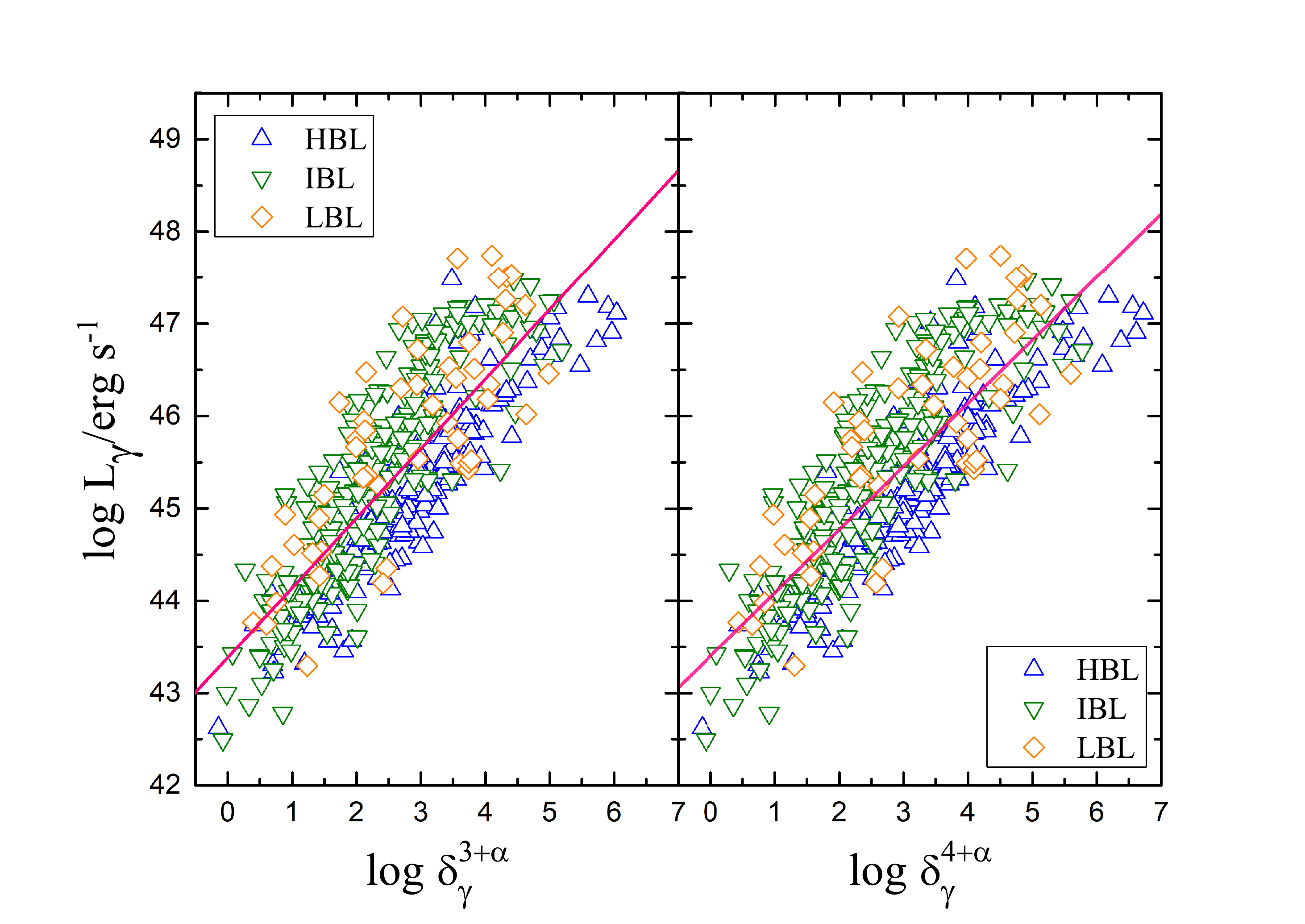}
   \caption{Plot of the correlation of $\log L_{\gamma}$ against $\log\delta_{\gamma}^{3+\alpha_{\gamma}}$ and $\log\delta_{\gamma}^{4+\alpha_{\gamma}}$. The best-fit relations of FSRQs signify that $\log L_{\gamma}=(0.71\pm0.03)\log\delta_{\gamma}^{3+\alpha_{\gamma}}+(43.91\pm0.13)$ and $\log L_{\gamma}=(0.62\pm0.02)\log\delta_{\gamma}^{4+\alpha_{\gamma}}+(43.96\pm0.12)$ (left panel). On the other hand, for BL Lacs, $\log L_{\gamma}=(0.75\pm0.02)\log\delta_{\gamma}^{3+\alpha_{\gamma}}+(43.38\pm0.07)$ and $\log L_{\gamma}=(0.68\pm0.02)\log\delta_{\gamma}^{4+\alpha_{\gamma}}+(43.40\pm0.07)$ (right panel), respectively.}
     \label{fig4}
\end{figure*}

Since the emission from the jet is strongly boosted by a factor $\delta^{p}$, thus one can assume that the $\gamma$-ray luminosity should be proportional to the $\gamma$-ray Doppler factor i.e., $L_{\gamma}\sim\delta_{\gamma}^{3+\alpha_{\gamma}}$ for stationary continuous jets or $L_{\gamma}\sim\delta_{\gamma}^{4+\alpha_{\gamma}}$ for moving sphere blob \citep{Fan13(b)}. We probe this relation in Figure \ref{fig4}. Within the case for continuous jets, we ascertain the best-fittings are,  
$$\log L_{\gamma}({\rm erg\,s^{-1}})=(0.71\pm0.03)\log\delta_{\gamma}^{3+\alpha_{\gamma}}+(43.91\pm0.13)$$ 
with $r=0.75$ and $P\sim0$ for FSRQ and   
$$\log L_{\gamma}({\rm erg\,s^{-1}})=(0.75\pm0.02)\log\delta_{\gamma}^{3+\alpha_{\gamma}}+(43.38\pm0.07)$$ 
with $r=0.82$ and $P\sim0$ for BL Lacs; 
besides, for a moving sphere, 
$$\log L_{\gamma}({\rm erg\,s^{-1}})=(0.62\pm0.03)\log\delta_{\gamma}^{4+\alpha_{\gamma}}+(43.96\pm0.12)$$
with $r=0.75$ and $P\sim0$ for FSRQ and 
$$\log L_{\gamma}({\rm erg\,s^{-1}})=(0.68\pm0.02)\log\delta_{\gamma}^{4+\alpha_{\gamma}}+(43.40\pm0.07)$$
with $r=0.82$ and $P\sim0$ for BL Lacs, are also obtained. The values of $\gamma$-ray Doppler factor in $\delta_{\gamma}^{3+\alpha_{\gamma}}$ and $\delta_{\gamma}^{4+\alpha_{\gamma}}$ from the relation above are derived by equation (\ref{eq11}) and (\ref{eq5}) respectively. We state parenthetically that although our derived $\delta_{\gamma}$ here is a lower limit, the above scaling relations should be valid for the cases when the estimated $\gamma$-ray Doppler factor $\ge$ this lower limit and also different variability timescales are taken into account.   

The tight relations verify that $L_{\gamma}\sim\delta_{\gamma}^{3+\alpha_{\gamma}}$ or $\sim\delta_{\gamma}^{4+\alpha_{\gamma}}$. The high-energy $\gamma$-rays detected from blazars implies that the presence of the beaming effect in those sources, otherwise, the $\gamma$-rays should have been absorbed due to pair production on collision with the lower-energy photons populating the region \citep{Mat93}. Thus from the point of view of the present considerations and our correlation analysis, the $\gamma$-ray emission is strongly beamed. It also manifests that we can utilise $\delta_{\gamma}$ to flourish our understanding of Doppler beaming effect. In addition, we can see that the slope corresponding to the continuous is 0.71 and 0.75 for FSRQs and BL Lacs respectively. This provides hints that the jet is possibly continuous in the $\gamma$-ray bands.    
 
\subsection{More implications}

The assumption we take for the estimation of $\delta_{\gamma}$, primarily basing upon that the X-ray emission is originated from the same region as $\gamma$-ray emission, and the intensities of X-ray and $\gamma$-ray are similar when $\gamma$-ray is observed. We do not have direct evidence that the $\gamma$-ray and the X-ray emission regions are cospatial is true in general, but have indirect evidence for some sources. For example, \citet{Pri20} presented a multiwavelength temporal and spectral analysis of flares of 3C 279 during the period from 2017 November to 2018 July, and three bright $\gamma$-ray flares were observed simultaneously in the X-ray and optical/UV band. A so-called ``harder-when-brighter'' tendency has been observed in both $\gamma$-rays and X-rays during the flaring period. The cross-correlation study of the emission from different wavebands has been performed and shown a strong correlation between them without any time lags. This is the first time that 3C 279 has shown a strong correlation between $\gamma$-ray and X-ray emission with zero time lag, indicating that their origin are cospatial. Besides, \citet{3C279} also reported the results of multiwavelength decade-long monitoring (starting from 2008 and ending at 2018) of 3C 279. The $\gamma$-ray and X-ray light curves correlate fairly with no delay $\gtrsim$ 3 hs, suggesting the co-spatiality of the $\gamma$-ray and X-ray emission regions. Recently, \citet{4C19} studied a high-redshift blazar 4C+71.07 from the multi-wavelength simultaneous observations, and one of those prominent results substantiated that the $\gamma$-ray emission and X-ray emission are perhaps produced within the same region.  

The blazar TXS 0506+056, i.e. 4FGL J0509.4+0542 \citep[$z=0.3365$,][]{Pai18}, the first detected neutrino emitter, was reported by the IceCube Neutrino Observatory on 2017 September 22 \citep{ICE18(b), ICE18(a)}, becoming a target of multiwavelength monitoring due to the detection of a 290 TeV muon-track neutrino event IceCube-170922A coincident with its direction and arrival time during a $\gamma$-ray flare, opening up the possibility of an association between blazars and very high energy (VHE) neutrinos, and providing the evidence for the blazar jet acting as an accelerator of cosmic-ray particles. \citet{Lio18} obtained the Doppler factor for TXS 0506+056 is 14.67 and \citet{Che18} reported 14.3. The above two values are quite close to the averages of their sample. However, \citet{Li20} has reported on the radio properties of the parsec-scale jet in TXS 0506+056 derived from the analysis of multiepoch and multifrequency archive very long baseline interferometry (VLBI) data. They obtained $\delta=2.65$. We derived the $\gamma$-ray Doppler factor for this source is 4.25 in this paper, with adopting $F_{\rm 1\,keV}=0.042$ $\mu$Jy, $E_{\gamma}=4.21$ GeV and $\alpha_{X}=1.89$. Our result is lower than that from \citet{Che18} or \citet{Lio18} but higher than that from \citet{Li20}. Considering the core-dominance parameter of this source to be $\log R=-0.13$ \citep{Pei20(a)}, and also take into account the fact that the estimated values of Doppler factor in our work and from other literature are not large, it suggests that a moderately relativistic jet originated in TXS 0506+056.           

Another neutrino event IceCube-190730A was reported to be in spatial coincidence with the bright $\gamma$-ray FSRQ PKS 1502+106, i.e. 4FGL J1504.4+1029, and the neutrino was reported with a signalness of 67\% and an energy of 300 TeV \citep{ICE18(b), Fra20}. This source is also included in our present sample and $\delta_{\gamma}=13.41$, which is much higher than the average value for our subsample of FSRQs. \citet{Che18} and \citet{Lio18} obtained a quite high Doppler factor for this source as well, 23.8 and 13.77 were reported, respectively. It shows that this neutrino emitter is strongly Doppler boosted. We calculated the $\gamma$-ray luminosity of $\log L_{\gamma}=48.49$ erg s$^{-1}$, which indicates that PKS 1502+106 is a very bright source in the 4FGL catalog in terms of $\gamma$-ray energy flux at >100 MeV even though it has a large redshift of 1.838 \citep{HW10}, indicating an extremely high intrinsic luminosity.

From the multiwavelength characterization of SED, \citet{Rig19} studied some neutrino blazars candidates with their physical composition and the emission processes, in which a HSP source RXJ1022.7-0112, i.e. 4FGL J1022.7-0112 is also listed in our sample. We report an extremely high lower limit on $\gamma$-ray Doppler factor for this source of $\delta_{\gamma}=21.22$. For another HSP source listing in both papers, TXS 0628-240, i.e. 4FGL J0630.9-2406, we obtained $\delta_{\gamma}=10.43$. \citet{Che18} also reported a very high Doppler factor for this neutrino candidate of 51.

In addition, we found some more blazars candidates of neutrino emitters in our present sample after cross-checking with the literature. For example, $\delta_{\gamma}=11.40$ for PKS B1424-418 i.e. 4FGL J1427.9-4206, which was in coincidence with the arrival time of a PeV neutrino \citep{Kad16, Fra20}; $\delta_{\gamma}=3.67$ for a distant BL Lac object MG3 J225517+2409 i.e. 4FGL J2255.2+2411 with a redshift of 1.37 \citep{Fra20, LAT19(a)}; $\delta_{\gamma}=6.48$ for FSRQ S4 1716+68 i.e. 4FGL J1716.1+6836, which the duration of the neutrino flare is short \citep{Fra20}; $\delta_{\gamma}=6.71$ for OJ 508 i.e. 4FGL J1022.7-0112 with statistically significant potential for neutrino emission \citep{Smi20}; $\delta_{\gamma}=5.19$ for HSP PMNJ0953-0840 i.e. 4FGL J0953.0-0840 \citep{Rig19}. Therefore, from our comparison, it reaches to an interesting finding that the lower limit on $\gamma$-ray Doppler factor for these neutrino emitter candidates are relatively quite high, suggesting that these sources are also possibly strongly Doppler boosted.

For the well-studied FSRQ 3C 279 (i.e. 4FGL J1256.1-0547), $\delta_{\gamma}=7.21$ are ascertained in this paper. \citet{Fan13(a), Fan14} also calculated the $\gamma$-ray Doppler factor for this source with adopting the data from 2FGL and obtained 5.99 and 5.62 respectively, both by setting $\Delta T$= 1 day. Our estimation here is higher than the results from \citet{Fan13(b)} or \citet{Fan14}, since some input parameters for calculation are quite different from there, $F_{\rm 1\,keV}=4.338$ $\mu$Jy, $E_{\gamma}=3.13$ GeV and $\alpha_{X}=0.84$ are adopted. \citet{Fan13(b)} and \citet{Fan14} both reported $E_{\gamma}=3.54$ which are similar with this paper but quite low $F_{\rm 1\,keV}$ were used, 1.340 $\mu$Jy and 0.961 $\mu$Jy presented in \citet{Fan13(b)} and \citet{Fan14}, respectively, and thus led their low estimation of $\delta_{\gamma}$. From constraining the brightness temperature, \citet{Lio18} found the Doppler factor of 11.64 for 3C 279, this value is lower than the average value of their sample. \citet{Che18} obtained $\delta=27.7$ that is higher than their average from SED modeling. Our derived result is also higher than our average value ($\langle\delta_{\gamma}\rangle\sim6.87$ for FSRQ). This could possibly suggest that the $\gamma$-ray and radio regions share the same relativistic effects.

The connection between $\gamma$-ray and parsec-scale radio flares has been claimed by previous studies \citep[e.g.][]{PKL10, Agu11(a), Jor13}. However, the location of $\gamma$-ray emission originating is still debated. Two scenarios have been proposed: on subparsec scales or on a few parsec scales. The former scenario indicates that the $\gamma$-ray emission zone is located close to the true base of the jet and the black hole \citep[e.g.][]{PKL10, Tav10, Ran14}. The latter one implies that $\gamma$-rays are produced from many parsecs away from the jet apex \citep[e.g.][]{Mar10, Agu11(b), Sch12}. If a $\gamma$-ray event occurs in a part of a relativistic jet that is optically thin for radio emission, then radio band variability is expected to be simultaneous at different radio frequencies and be simultaneous or even precede the $\gamma$-ray variability \citep[e.g.][]{GT08, Agu11(a), Jor13}.

\citet{Lis17} performed a comprehensive VLBA study of the blazar 3C 273 associated with a strong $\gamma$-ray flare, and found that the $\gamma$-ray emission region in 3C 273 is close to the true base of the jet, located $2-7$ pc upstream from the 7 mm radio core, implying that the location of $\gamma$-ray emission and radio emission is in the same jet in 3C 273 but different site. 

Relativistic beaming in blazars is evidenced by the rapid variability property from radio to $\gamma$-ray wavelengths \citep{All96, LV03, Kov09}, and it is originated from within the AGN jet and strongly beamed along the line of sight. This yields an interesting fact that all $\gamma$-ray-loud AGNs are also radio-loud while not all radio-loud AGNs are $\gamma$-ray loud. This could be explained by two ways: the $\gamma$-ray emission is in a Doppler boosted state inside a narrow conoid if the $\gamma$-ray emitting regions are moving faster than the radio emitting regions \citep{SS94}. Then outside the $\gamma$-ray beaming conoid and within the radio beaming conoid, the $\gamma$-ray radiation would be in a Doppler dimmed state, displaying that $\gamma$-ray-quiet but radio-loud AGNs.

Another scenario for explicating this fact should be related to the AGN jet bending \citep{Mon95}. Figure \ref{fig2} shows that around one third (left panel) and one fourth of sources (right panel) locate at the upper left of the equality line, implying that the derived lower limit on $\delta_{\gamma}$ for those sources are higher than that from the radio estimation. This could be possibly explained by the jet bending within those blazars. If a bend in the jet takes place at upstream of the $\gamma$-ray emission region and downstream of extended radio emitting region, then the radiation would be Doppler boosted in alignment to their respective sections of the jet. If so, the $\gamma$-ray emission is beamed toward our line of sight (resulting in a high $\delta_{\gamma}$) while the radio emission is beamed away from us (resulting in a low $\delta_{\gamma}$) ascribable to a better alignment of the $\gamma$-ray emitting jet section with our line of sight. In this case, we would probably detect a $\gamma$-ray-loud and radio-loud AGNs.

\section{CONCLUSION}

From the beaming model we can expect that the more core-dominated sources should be more beamed and have larger Doppler boosting factors consequently. In this paper, we implement a method to calculate the lower limit on $\gamma$-ray Doppler factors ($\delta_{\gamma}$) for total 809 $\gamma$-ray blazars detected by {\it Fermi}, including 342 FSRQs and 467 BL Lacs (202 HBLs, 213 IBLs and 52 LBLs), and also study the relation between the core-dominance parameter ($R$) and $\delta_{\gamma}$ for 507 sources, suggesting that the core-dominance parameter is perhaps an indicator for beaming effect. We use the derived results of $\delta_{\gamma}$ to probe the intrinsic property of $\gamma$-ray Doppler boosted effect and shed new light on $\gamma$-ray blazars. Even though there are indeed some discrepancies between two sorts of Doppler factors basing on different assumptions (even with regarding to other methods), the nature perhaps originates from the same population. The main conclusions of this work are the following:

\begin{itemize}
  \item The average value of the lower limit on $\gamma$-ray Doppler factor for FSRQs and BL Lacs are obtained, $\langle\delta_{\gamma}|_{\rm FSRQ}\rangle=6.87\pm4.07$ and $\langle\delta_{\gamma}|_{\rm BL\,Lac}\rangle=4.31\pm2.97$ respectively, suggesting that the $\gamma$-ray emission of blazars is strongly beamed. 
  \item The $\gamma$-ray Doppler factor closely correlates with the variability Doppler factor from our comparison, suggesting that the $\gamma$-ray and radio regions possibly share the same relativistic effects.
   \item The lower limit on $\delta_{\gamma}$ for some sources are higher than that from the radio estimation, which is believed to be due to the jet bending in those blazars.
   \item From our analysis, we obtain an interesting finding that the on average variability timescale for {\it Fermi}-detected blazars is possibly around 1.5$\sim$2 hs. 
   \item The correlation between $\delta_{\gamma}$ and $\gamma$-ray luminosity suggests the jet is possibly continuous in the $\gamma$-ray bands. 
   \item We predict that the blazars candidates of neutrino emitters are potentially strongly Doppler boosting sources. 
   \item The $\gamma$-ray Doppler factor is correlated with the core-dominance parameter, indicating that $R$ can be taken as the indicator for relativistic beaming effect.
    
\end{itemize}

\begin{acknowledgements}
We greatly thank the anonymous referee for valuable comments and suggestions, which help us to improve the manuscript. The work is partially supported by the National Natural Science Foundation of  China ( NSFC 11733001, NSFC U1531245), Natural Science Foundation of Guangdong Province (2019B030302001; 2017A030313011), supports for Astrophysics Key Subjects of Guangdong Province and Guangzhou City.
\end{acknowledgements}

\bibliographystyle{pasa-mnras}
\bibliography{Pei}

\end{document}